\documentclass[%
 superscriptaddress,
 amsmath,amssymb,
 aps,
]{revtex4-1}

\usepackage{graphicx}
\usepackage{dcolumn}
\usepackage{bm}
\usepackage{xcolor}
\usepackage{url}
\usepackage{float}
\usepackage{array}

\usepackage{booktabs}

\usepackage{subfigure}
\usepackage{multirow}

\begin{document}


\title{Determination of the Gluonic D-term and Mechanical Radii of Proton from Experimental data}
\author{Wei Kou}
\email{kouwei@impcas.ac.cn}
\affiliation{Institute of Modern Physics, Chinese Academy of Sciences, Lanzhou 730000, China}
\affiliation{University of Chinese Academy of Sciences, Beijing 100049, China}

\author{Rong Wang}
\email{rwang@impcas.ac.cn}
\affiliation{Institute of Modern Physics, Chinese Academy of Sciences, Lanzhou 730000, China}
\affiliation{University of Chinese Academy of Sciences, Beijing 100049, China}
\author{Xurong Chen}
\email{xchen@impcas.ac.cn}
\affiliation{Institute of Modern Physics, Chinese Academy of Sciences, Lanzhou 730000, China}
\affiliation{University of Chinese Academy of Sciences, Beijing 100049, China}
\affiliation{Guangdong Provincial Key Laboratory of Nuclear Science, Institute of Quantum Matter, South China Normal University, Guangzhou 510006, China}



\begin{abstract}
We apply a ``color" tripole ansatz for describing the $D$-term of the proton.
By fitting the experimental data of the vector meson J/$\psi$ and $\phi$ photoproductions near the thresholds,
we firstly obtained the gluonic $D$-term of the proton.
$D_g(0)$ is estimated to be $-2.16\pm0.42$ for J/$\psi$ and $-1.31\pm0.48$ for $\phi$,
and the mechanical root mean square radius of proton is estimated to be $0.61\pm0.29$ fm for $\phi$ and $0.42\pm0.11$ fm for J/$\psi$. 
\end{abstract}

\maketitle


\section{Introduction}
\label{sec:intro}

Recently in the study of hadron physics, the mechanical properties and
the quantum chromodynamics (QCD) \cite{Gross:1973id,Gross:1973ju,Gross:1974cs} structure of the proton
are among the hot topics under discussions.
The proton is composed of quarks and gluons according to
the quark model and the modern QCD theory.
Although one can break the atomic nucleus to get the nucleons inside,
we cannot observe the individual quarks and gluons by breaking up the nucleons.
This is because of the color confinement of QCD theory.
Nevertheless the proton structure can be revealed
in the asymptotic region under high energy transfer.
There are some theoretical ways to envisage the internal structure of the proton.
The QCD factorization theory indicates that the scattering amplitudes
can be expressed in terms of the generalized parton distribution (GPD) \cite{Ji:1996ek,Mueller:1998fv,Radyushkin:1996nd}
in the high-$Q^2$ region. Although less rigorous, there are some phenomenal models
which also successfully explain parts of physical processes.

The basic mechanical properties of proton can be deduced from
the gravitational form factors (GFFs) of Energy-Momentum-Tensor (EMT) \cite{Ji:1996nm,Pagels:1966zza,Belitsky:2005qn}.
Matrix elements of EMT contain basic mechanical properties such as mass and angular momentum.
Over the past two decades people have gradually understood the origins of the proton mass and spin
within the framework of quantum field theory and from the prior experimental studies.
But much less is known about the $D$-term of the proton,
which is one of the GFFs that reflects the shear force and the pressure inside the proton \cite{Polyakov:2002yz,Polyakov:2018zvc}.
The old fashion models on the $D$-term were performed in the bag model \cite{Ji:1997gm}
and the chiral quark soliton model ($\chi$QSM) \cite{Petrov:1998kf}.
In 2018 V. Burkert, et. al. \cite{Burkert:2018bqq} adopted GPD measurement
from deeply virtual Compton scattering (DVCS) \cite{Ji:1996nm,Ji:1996ek}
to firstly give a direct mapping of the pressure distribution
inside proton with the extracted $D$-term $D_q(0)$ of quark.
In the following year, P. Shanahan and W. Detmold gave the gluonic and total $D$-term
by Lattice QCD (LQCD) calculation \cite{Shanahan:2018pib,Shanahan:2018nnv}.
Hatta et. al. have analyzed the $D$-term
by studying the photoproduction process of heavy quarkomium
from the holographic QCD with Operator Product Expansion theory \cite{Hatta:2018ina,Boussarie:2020vmu,Hatta:2021can}.
H. Dutrieux et. al. apply the artificial neural network technique
recently to analyze the DVCS process in extracting the $D$-term of the proton \cite{Dutrieux:2021nlz}.
	
On experimental side, the extraction of the gluonic $D$-term is rarely mentioned in the literatures.
The exclusive leptoproduction of vector mesons $e p \rightarrow e^{\prime} \gamma^{*} p \rightarrow e^{\prime} p^{\prime} V$
can broadly unveil the details of the structure of proton.
The proton has only the up and down valence components,
and the light mesons can be used to detect only these partons in proton.
But in the Regge limit, the heavy quarkonium like J/$\psi\ \mathrm{and}\ \Upsilon$
can be used to probe the gluon part of the proton.
It is similar to the method of extracting the gravitational radius of the proton
from the GFFs of EMT \cite{Kharzeev:2021qkd,Wang:2021dis}.
In this work, we apply the vector meson $\phi,\ \mathrm{J}/\psi$ near-threshold photoproduction data \cite{Mibe:2005er,Ali:2019lzf}
to extract the gluons' contribution of $D$-term $D_{g}(0)$.
Then we present the internal pressure shear force distributions of the proton with the obtained $D$-term.

The organization of the paper is as follows.
The EMT and $D$-term of the proton are briefly introduced in Sec. \ref{sec:EMT-and-D-term}.
The result on the parameterized $D$-term extracted from vector meson photoproduction is present in Sec. \ref{sec:Extraction-of-D-term}.
The pressure and shear forces distribution inside the proton is shown in Sec. \ref{sec:Pressure}. We also get the $D$-term radius defined by Zahed's work \cite{Mamo:2021krl} and mechanical radius defined by \cite{Polyakov:2018zvc} in Sec. \ref{sec:radii}. 
A summary is given in Sec. \ref{sec:summary}.

\section{Energy-momentum tensor and $\bf D$-term}
\label{sec:EMT-and-D-term}

The QCD EMT is the conserved current associated with the space-time translational symmetry based on Noether's theorem.
The total tensor satisfies the conservation law:
	\begin{equation}
		\begin{aligned}
		&\partial^{\mu} \hat{T}_{\mu \nu}=0, \\
		&\hat{T}_{\mu \nu}=\sum_{q} \hat{T}_{\mu \nu}^{q}+\hat{T}_{\mu \nu}^{g}.
		\end{aligned}
		\label{eq:conserve-eq}
	\end{equation}
The covariant normalization condition is set as
$\left\langle p^{\prime} \mid p\right\rangle=2 p^{0}(2 \pi)^{3} \delta^{(3)}\left(\boldsymbol{p}^{\prime}-\boldsymbol{p}\right)$,
where the state $|\boldsymbol{p}\rangle$ is one particle state.
The form factors of EMT depend the kinematic variables $P=\frac{1}{2}\left(p^{\prime}+p\right), \Delta=p^{\prime}-p, t=\Delta^{2}$.
As the proton is the focus of this paper, the matrix element of the QCD EMT of the Spin-$\frac{1}{2}$ hadron is defined as \cite{Polyakov:2018zvc},
	\begin{equation}
		\begin{aligned}
			&\left\langle p^{\prime}, s^{\prime}\left|\hat{T}_{\mu \nu}^{a}(x)\right| p, s\right\rangle=\bar{u}^{\prime}\left[A^{a}(t) \frac{\gamma_{\{\mu} P_{\nu\}}}{2}\right.\\
			&\left.+B^{a}(t) \frac{i P_{\{\mu} \sigma_{\nu\} \rho} \Delta^{\rho}}{4 m}\right.\left.+D^{a}(t) \frac{\Delta_{\mu} \Delta_{\nu}-g_{\mu \nu} \Delta^{2}}{4 m}\right.\\
			&+m \bar{c}^{a}(t) g_{\mu \nu}\bigg] u e^{i\left(p^{\prime}-p\right) x},
		\end{aligned}
	\label{eq:matrix-ele}
	\end{equation}
where the normalization of the spinor is $\bar{u}(p, s) u(p, s)=2 m$,
with the notation $a_{\{\mu} b_{\nu\}}=a_{\mu} b_{\nu}+a_{\nu} b_{\mu}$.
The individual quark and gluon form factors $A^{a}(t),\ B^{a}(t),\ D^{a}(t)$
and $\bar{c}^{a}(t)$ depend on the renormalization scale \cite{Polyakov:2018zvc}.
From the lattice QCD calculation \cite{Alexandrou:2017oeh}, the $B^{a}(t)$ is suggested to be zero for the gluon.

The other GFFs besides the $D$-term form factor contain the information of the mass and the spin distributions of the proton.
In this paper we mainly focus on the $D$-term form factors which can describe the coupling between the vector meson and the proton.
The GFFs related to the vector production is discussed in Ref. \cite{Hatta:2021can}.
The tripole parametrization ansatz for the $D$-term is
suggested by the perturbative counting rule at large $t$ \cite{Tanaka:2018wea,Hatta:2021can},
	\begin{equation}
		 D_{q,g}(t)=\frac{D_{q,g}(0)}{\left(1-t / m_{D}^{2}\right)^{3}}.
		\label{eq:D-term}
	\end{equation}
The tripole form contributes more to the scattering amplitude in the small $t$ region.

\section{Extraction of $\bf {D_g(0)}$}
\label{sec:Extraction-of-D-term}

Usually both GFF $A(t)$ and GFF $D(t)$ couple to the vector meson photoproduction.
$D$-term dominates at $t$ approaching zero and it has the steeper $t$-dependence.
To extract the $D$-term from the vector meson near-threshold photoproduction data,
we assume that \cite{Hatta:2018ina},
	\begin{equation}
		\frac{d\sigma}{dt}\sim D^{2}(t)=\frac{D_{g}^2(0)}{\left(1-t / m_{D}^{2}\right)^{6}}.
		\label{eq:cross-section}
	\end{equation}
Since the form factors are evaluated at a large scale $\mu^2=Q^2$,
in an approximation we can use the following asymptotic results,
	\begin{equation}
		\begin{aligned}
		A_{q}(0) &\approx \frac{n_{f}}{4 C_{F}+n_{f}},\ 
		 \quad A_{g}(0) &\approx \frac{4 C_{F}}{4 C_{F}+n_{f}},\ 
		  \quad D_{q}(0) &\approx \frac{n_{f}}{4 C_{F}} D_{g}(0),
		\end{aligned}
	\label{eq:asymptotic}
	\end{equation}
where $C_{F}=\frac{N_{c}^{2}-1}{2 N_{c}}$ is the Casimir operator with $N_c=3$
and $n_f=3$ represents the number of light flavors in the proton.
The aim of the analysis is to extract $D(0)$, therefore we use the third row in Eq. (\ref{eq:asymptotic}),
which relates the quark $D$-term and the gluon $D$-term with $D_q(0)=\frac{9}{16}D_g(0)$.
In analyzing the experimental data of the photoproductions of $\phi$ and J/$\psi$,
we assume that only the gluon component of the proton are involved in the color interaction near the threshold.
This is reasonable given that s and c quarks in the
nucleon are generated by the gluon splitting $g\to s\bar{s} / c\bar{c}$ \cite{Frankfurt:2002ka}.
The gluon $D(0)$ is directly regarded as the total one, as $D_g(0)=D(0)$, because of the weak gravitational field approximation \cite{Kharzeev:2021qkd}.
For $\phi$ and J/$\psi$ meson productions, Eq. (\ref{eq:cross-section}) is used to
fit the experimental data \cite{Mibe:2005er,Ali:2019lzf}.
The cutoff parameter $m_D$ is also a free parameter to be determined by the fits. 

\begin{figure}[H]
	\centering  
	\subfigure[]{
		\label{fig:phi-fit}
		\includegraphics[width=0.45\textwidth]{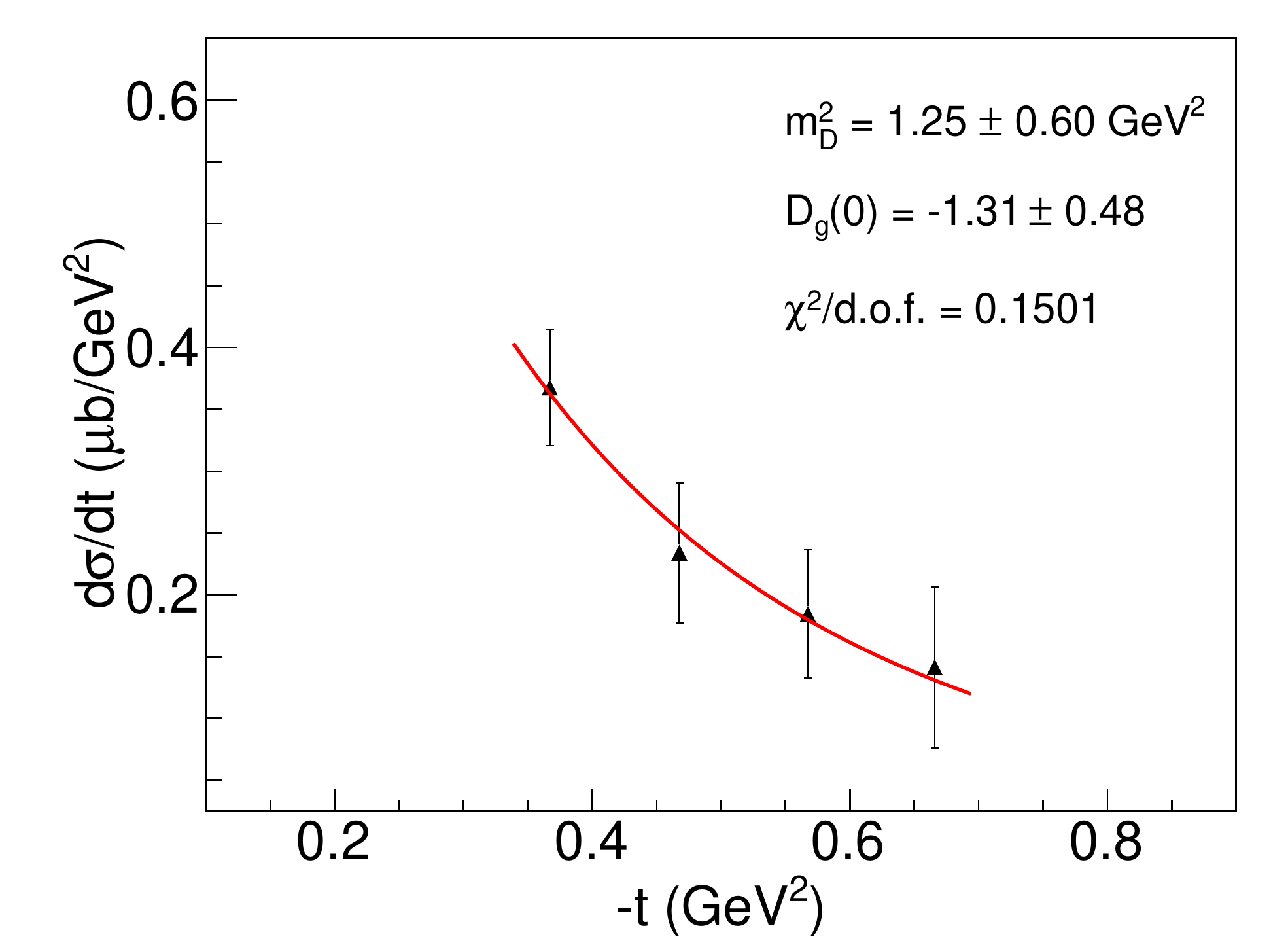}}
	\subfigure[]{
		\label{fig:psi-fit}
		\includegraphics[width=0.45\textwidth]{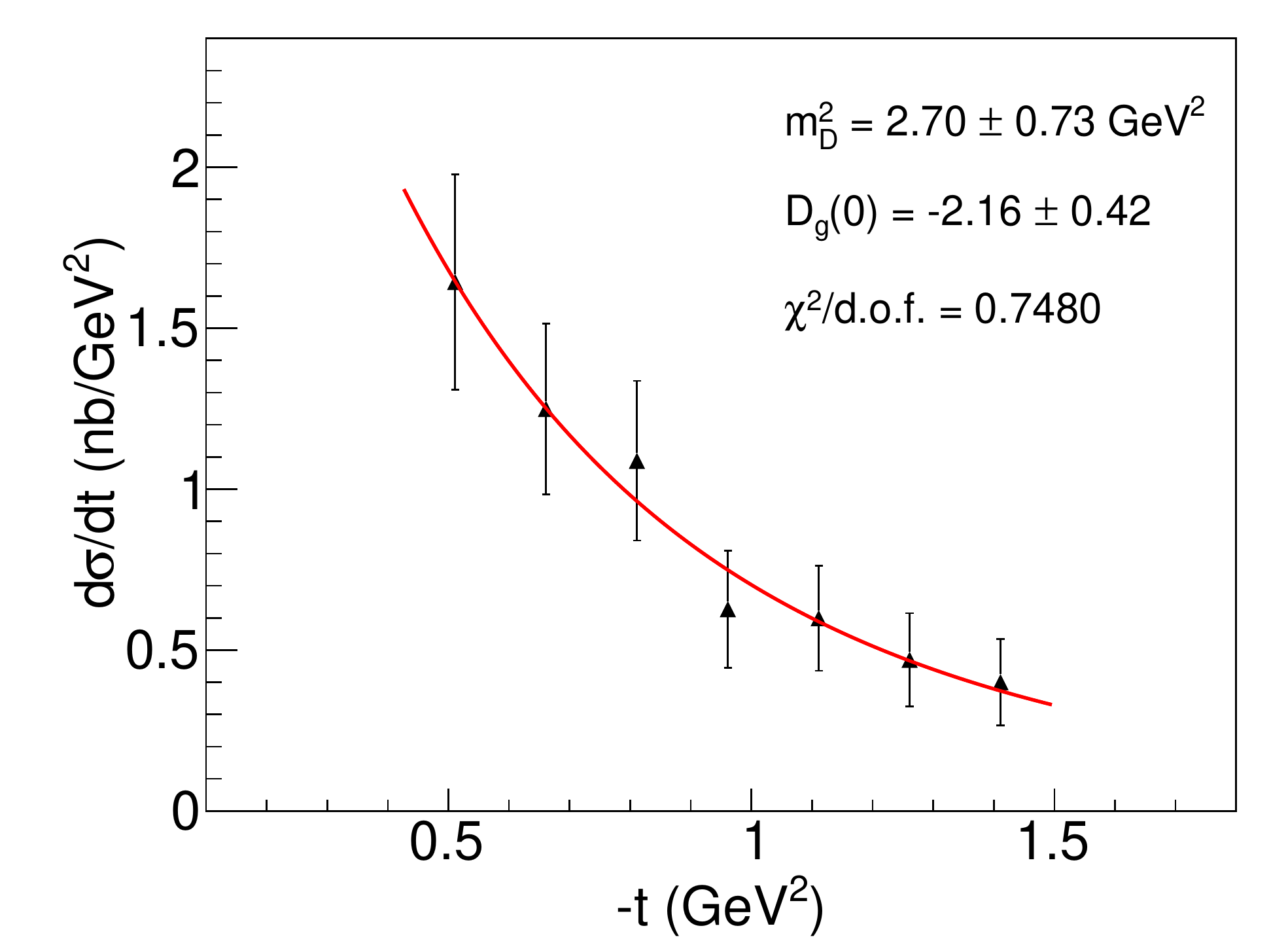}}
	\caption{(Color online) (a) The differential cross section of $\phi$ photoproduction near threshold
		at $E_{\gamma}=$ 1.62 GeV by LEPS Collaboration \cite{Mibe:2005er}.
		The red curve is our fit with the tripole form, with $m_D^2=1.25\pm0.60$ GeV and $D_g(0)=-1.31\pm0.48$. (b) The result by \cite{Ali:2019lzf} which have the similar marks by (a). }
	\label{fig:fitting}
\end{figure}

\begin{figure}[htbp]
		\centering
		\includegraphics[width=0.42\textwidth]{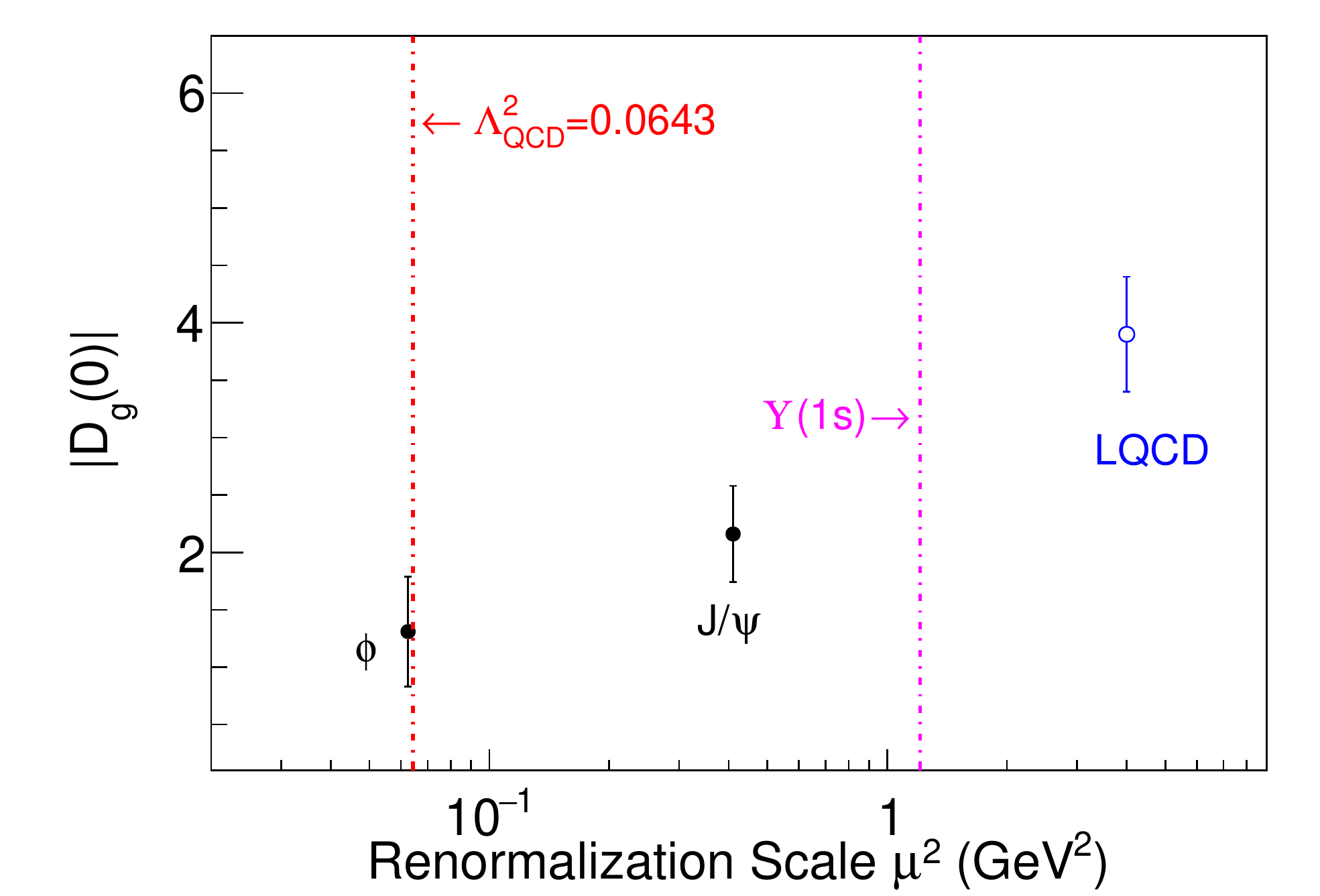}
		\caption{(Color online) The gluon contribution to the $D$-term inside proton extracted from the experimental data (black dots)
                  and the LQCD calculation \cite{Shanahan:2018pib} (blue circle).
                  The renormalization scale is indicated with the horizontal axis,
                  which is defined as the the binding energy of the vector meson.
                  The scale of LQCD calculation corresponds the $\overline{\mathrm{MS}}$ scheme at a scale of $\mu=2$ GeV.
                  The magenta doted line marks the binding energy of $\Upsilon$ meson. }
		\label{fig:compare}
	\end{figure}

The near-threshold photoproduction data of $\phi$ and J/$\psi$ vector mesons from the LEPS \cite{Mibe:2005er} and GlueX \cite{Ali:2019lzf} Collaborations
with the fits of the tripole form $D$-term
are shown in FIG. \ref{fig:phi-fit} and FIG. \ref{fig:psi-fit}, respectively.
Note that only the gluon contribution is considered.
The obtained values of the $D$-term are shown in FIG. \ref{fig:compare},
compared with the result from LQCD \cite{Shanahan:2018pib}.
The extracted parameters and the qualities of the fits are summarized in Table \ref{tab:information}.
The results of $\Upsilon$ will be verified with the future high-precision experiments.

\section{Pressure and shear force distribution inside the proton}
\label{sec:Pressure}

\begin{figure}[H]
	\centering  
	\subfigure[]{
		\label{fig:pressure-phi}
		\includegraphics[width=0.45\textwidth]{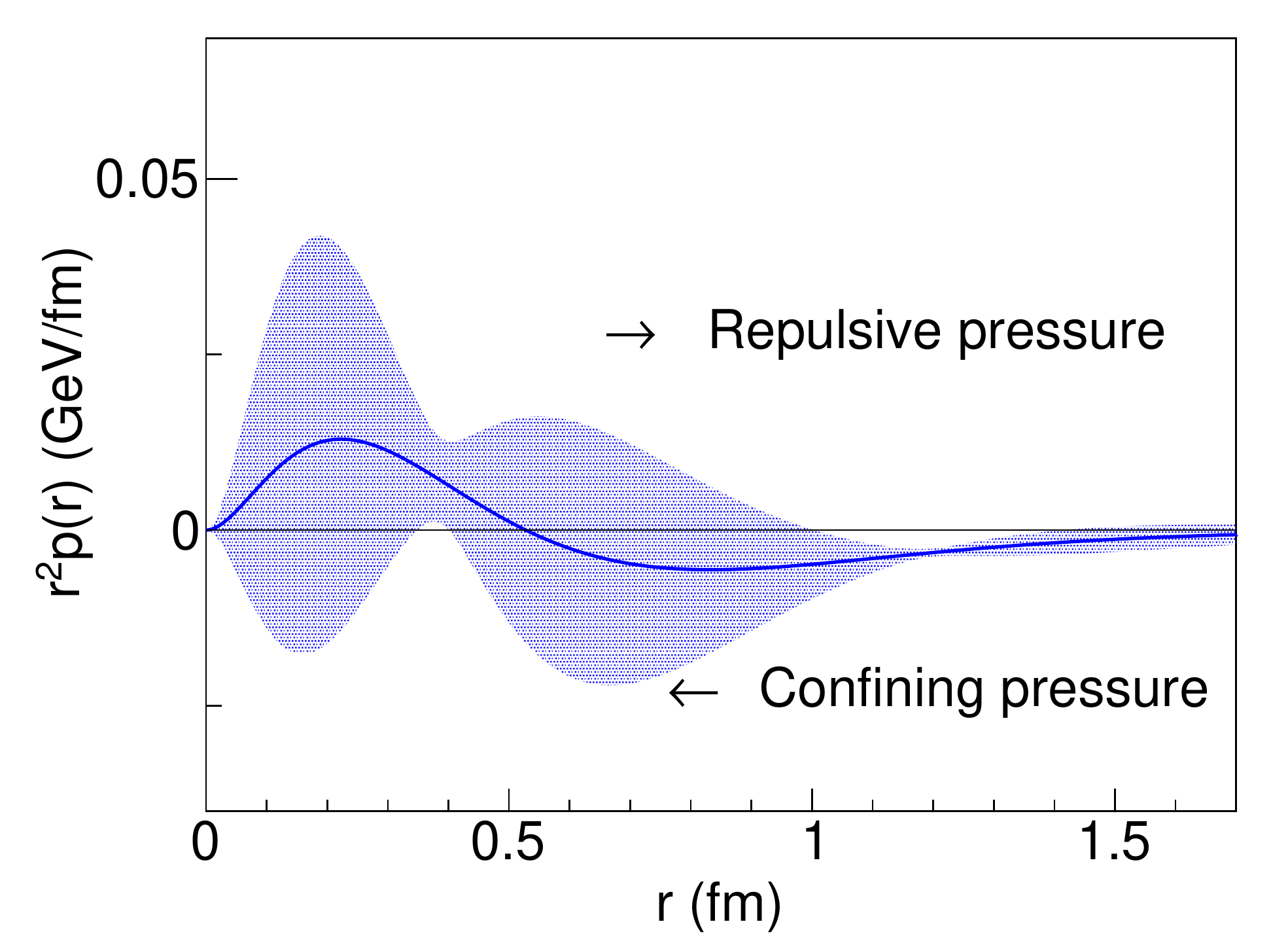}}
	\subfigure[]{
		\label{fig:pressure-jpsi}
		\includegraphics[width=0.45\textwidth]{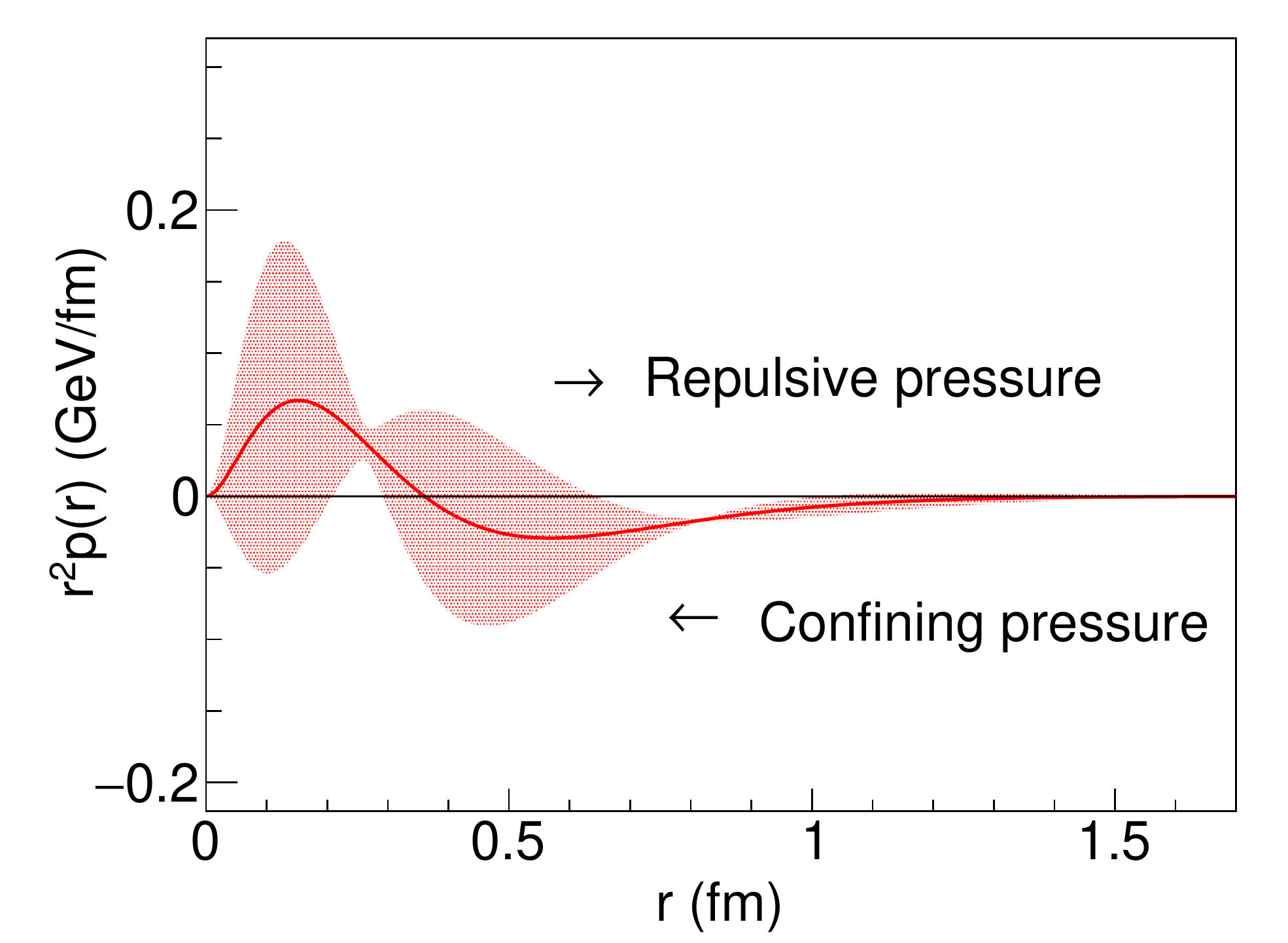}}
	\caption{(Color online) (a) The radial pressure distribution inside the proton based on
		the $D$-term extracted from the $\phi$ photoproduction.
		$r^2p(r)$ is the pressure distribution from the interactions of the gluons
		and $r$ is the radial distance to the center of the proton.
		The solid curve shows the center value of the pressure and band represents the statistical uncertainty. (b) The radial pressure distribution inside the proton based on
		the $D$-term extracted from the J/$\psi$ photoproduction. The description as same as (a). }
	\label{fig:pressure}
\end{figure}

\begin{figure}[H]
	\centering  
	\subfigure[]{
		\label{fig:shear-phi}
		\includegraphics[width=0.45\textwidth]{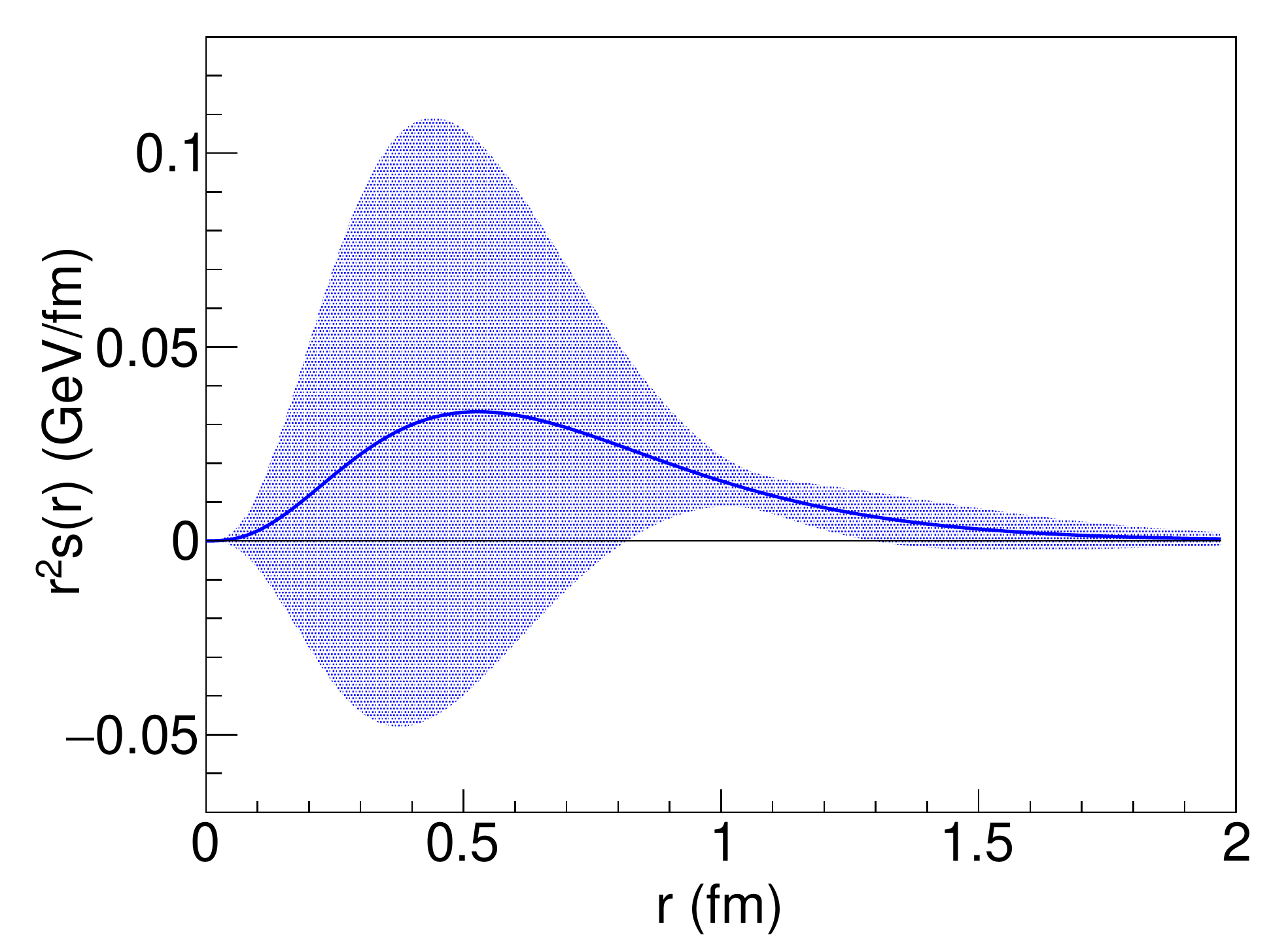}}
	\subfigure[]{
		\label{fig:shear-jpsi}
		\includegraphics[width=0.45\textwidth]{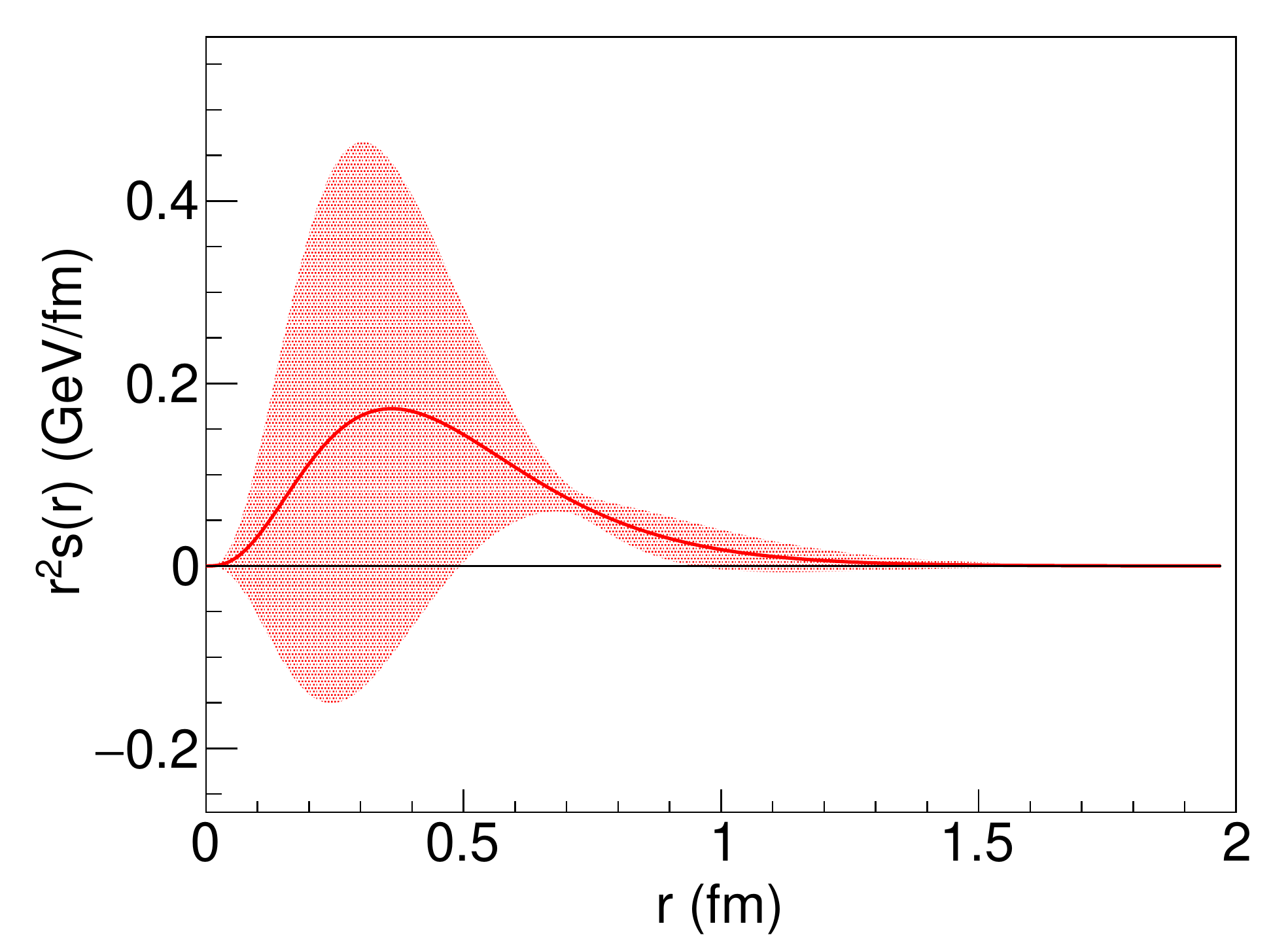}}
	\caption{(Color online) (a) The shear force distribution inside the proton based on
		the $D$-term extracted from the $\phi$ photoproduction.
		$r^2s(r)$ is the pressure distribution from the interactions of the gluons
		and $r$ is the radial distance to the center of the proton.
		The solid curve shows the center value of the pressure and band represents the statistical uncertainty. (b) The shear force distribution inside the proton based on
		the $D$-term extracted from the J/$\psi$ photoproduction. The description as same as (a). }
	\label{fig:shear}
\end{figure}

In the previous section, the values of $D_g(0)$ are given by analyzing the experimental data.
Generally, $D(t)$ contains the information of the internal pressure and shear force distributions
of the proton. Using the tripole parameterization for $D_g(t)$ and $-t=K^2$, we get,
\begin{equation}
	D_g(K)=\frac{D_g(0)}{\left(1+\frac{K^{2}}{m_{D}^{2}}\right)^{3}}.
	\label{eq:momentum-space}
\end{equation}
The Fourier transform of Eq. (\ref{eq:momentum-space}) gives the $D$-term in the three-dimensional coordinate space ($E = m_N$), as,
\begin{equation}
	\begin{aligned}
	\tilde{D}_g(r)=D_g(0) \int \frac{d^{3} K}{2 E(2 \pi)^{3}} \frac{e^{-i K \cdot r}}{\left(1+\frac{K^{2}}{m_{D}^{2}}\right)^{3}}
	=\frac{D_g(0)}{64\pi m_N}\left(\frac{1}{m_D }+r\right)m_D^4\exp(-m_Dr).
	\end{aligned}
\label{eq:Fourier}
\end{equation}
The pressure distribution $p(r)$ and shear force $s(r)$ inside the proton then can be computed with $\tilde{D}_g(r)$ \cite{Polyakov:2018zvc}:
\begin{equation}
	\begin{aligned}
	p(r)&=\frac{1}{3} \frac{1}{r^{2}} \frac{d}{d r} r^{2} \frac{d}{d r} \tilde{D}_g(r)=\frac{D_g(0)}{192\pi m_N}(m_Dr-3)m_D^5\exp(-m_Dr),\\
	s(r)&=-\frac{1}{2}\frac{d}{dr}\frac{1}{r}\frac{d}{dr} \tilde{D}_g(r)=-\frac{D_g(0)}{128\pi m_N}m_D^5r\exp(-m_Dr).
	\end{aligned}
	\label{eq:pressure}
\end{equation}
According to Eq. (\ref{eq:Fourier}--\ref{eq:pressure}), we calculate the pressure and shear force distributions inside the proton
with the associated uncertainties.
The uncertainties come from the uncertainties of the parameters $D_g(0)$ and $m_D$.
Using Eq. (\ref{eq:pressure}) the pressure and shear force distributions inside the proton are obtained and
displayed in FIG. \ref{fig:pressure-phi} to FIG. \ref{fig:shear-jpsi}.

As shown in this analysis,
the $D_g(t)$ extracted from the exclusive $\phi$ and J/$\psi$ production channels
are different, which results in the different pressure and shear force distributions as well.
According to our uncertainty analysis,
the uncertainty of the pressure and shear force distributions largely
depend on the error of parameter $m_D$ from the fit.
Constraining better the slope parameter $m_D$
would better to give the precise pressure and shear force distributions inside the proton.

\section{Root Mean square radii of proton}
\label{sec:radii}
\begin{figure}[htbp]
	\centering
	\includegraphics[width=0.42\textwidth]{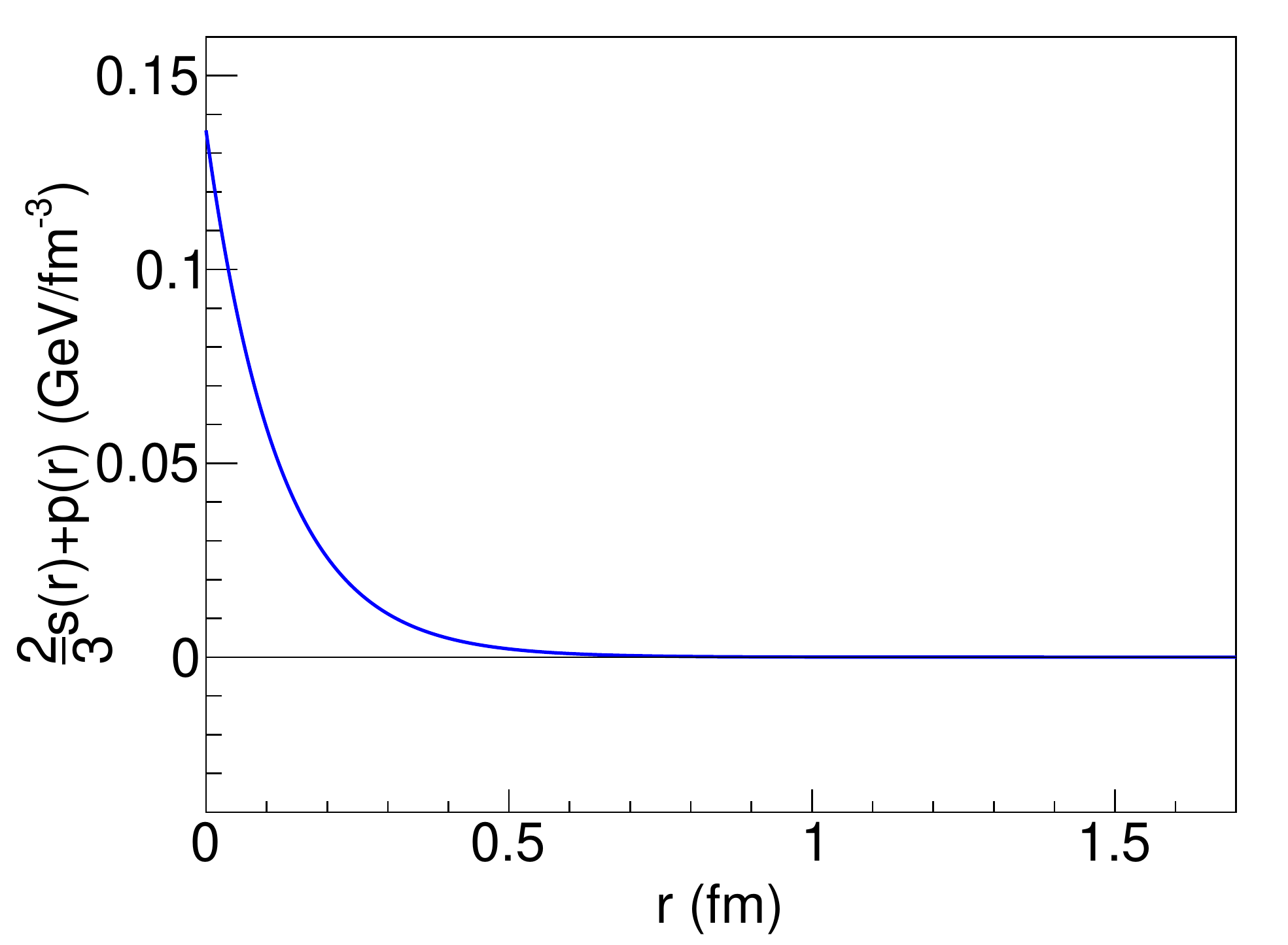}
	\caption{(Color online) The positive combination satisfy the relation from Eq. (\ref{eq:inequality}) using experimental data by J/$\psi$ production.}
	\label{fig:positive}
\end{figure}

According to the definition of root mean square radius \cite{Polyakov:2018zvc}, we can get the corresponding radius of $D$-term. To compare our results, we now introduce the mechanical force there. In previous section, we use the experimental data to have the fitting parameters and the pressure and shear force distributions $p(r)$ and $s(r)$. According the argument from \cite{Perevalova:2016dln}, authors thought that for the mechanical stability of the system the corresponding force must be directed outwards. Therefore the local criterion for the mechanical stability can be formulated as the inequality,
\begin{equation}
\frac{2}{3}s(r)+p(r)>0.
\label{eq:inequality}
\end{equation}
Of course, our results satisfy the inequality. We also use the parameters from fitting data of J/$\psi$ to give the figure as FIG. \ref{fig:positive}. 

The positive combination $\frac{2}{3}s(r)+p(r)$ is described as the meaning of the normal force distribution in the proton system. The mechanical radius for proton is introduced as \cite{Polyakov:2018zvc}
\begin{equation}
\left\langle r^{2}\right\rangle_{\mathrm{mec}}=\frac{\int d^{3} r r^{2}\left[\frac{2}{3} s(r)+p(r)\right]}{\int d^{3} r\left[\frac{2}{3} s(r)+p(r)\right]}=\frac{6 D_g(0)}{\int_{-\infty}^{0} d t D_g(t)}.
\label{eq.mechanical radius}
\end{equation}
For the tripole ansatz like Eq. (\ref{eq:D-term}) the resulting mechanical mean square radius is $\langle r^2\rangle_{\mathrm{mech}}=12/m_D^2$. To compare with the $D$-term radius, we use the classical radius definition like previous works \cite{Adamuscin:2012zz,Song:2018fhh,Kumano:2019vlv,Wang:2021dis}
\begin{equation}
\left\langle r^{2}\right\rangle_{\mathrm{D}}=\frac{6}{D_g(0)} \frac{dD_g(t)}{dt}\Bigg|_{t=0}=\frac{18}{m_D^2}.
\label{eq:d-term radius}
\end{equation}
\begin{table}[H]
	\caption{The least-square fit results to the experimental data \cite{Mibe:2005er,Ali:2019lzf} using Eq. (\ref{eq:cross-section})
		and the obtained parameters, mechanical radii and $D$-term radii for the tripole parametrization. The last line come from \cite{Burkert:2021ith} and we calculate the radii corresponding the DVCS data.
	}
	\begin{center}
		
		\begin{tabular}{ |c|c|c|c|c| }
			\toprule[1.3pt]
			Data    &  $D_g(0)$ & $m_D^2$ (GeV$^2$\textsc) &$\sqrt{\langle r^{2}\rangle_{\mathrm{mec}}}$ (fm) &$\sqrt{\langle r^{2}\rangle_{\mathrm{D}}}$ (fm) \\
			\hline
			$\phi$& $-1.31\pm0.48$&$1.25\pm0.60$&$0.61\pm0.29$&$0.75 \pm0.36
			$\\
			\hline
			J/$\psi$& $-2.16\pm0.42$ &$2.70\pm0.73$&$0.42\pm0.11$&$0.51\pm0.14$\\
			\hline
			DVCS \cite{Burkert:2021ith} &$-1.47\pm0.06 $ &$1.02 \pm0.13$ &$0.68\pm0.09$ &$0.83\pm0.10$\\
			\bottomrule[1.3pt]
		\end{tabular}	
	\end{center}
	\label{tab:information}
\end{table}

Table \ref{tab:information} concludes the different radii of proton using Eq. (\ref{eq.mechanical radius}) and (\ref{eq:d-term radius}). It can be clearly seen from these results that the mechanical radii are obviously smaller than the $D$-term radii. Both sets of experimental data illustrate this conclusion.
\section{Discussions and Summary}
\label{sec:summary}

One important outcome of the QCD factorization theorem is 
that for the process dominated by the two-gluon
exchange, the $t$-dependence should reach a universal limit which is independent
of the flavor of the quark constituents of the meson \cite{Brodsky:1994kf,Frankfurt:2002ka},
at large $Q^2$ and fixed $x$.
The mechanism for such universality is the transverse squeezing
of the meson wave function. From LQCD calculation \cite{Shanahan:2018pib,Shanahan:2018nnv},
the authors concluded that the gluons play an important role
in the internal dynamics of the proton, which is quite distinct from that of quarks.  
Hence in this limit the $t$-dependence of the amplitude is given solely by the
gluons' contribution.
Therefore it is a reasonable assumption to extract the gluonic GFFs
from the differential cross section data of the vector meson photoproductions
in the region close to the threshold energy.

In this work, the gluons' contribution to the proton $D$-term
and the pressure distribution inside the proton are determined 
from the vector meson photoproduction experiments near the threshold.
The vector meson mass dependence of the gluonic $D$-term is
shown in FIG. \ref{fig:compare}.
Our speculation for the dependence is that
the gluonic $D$-term is scale-dependent
and different meson probes correspond to different energy scales.
We note that a recent work presents a determination of the shear force distribution in the proton
using a DVCS experiment which measures the GPDs of the proton \cite{Burkert:2021ith}.
Our work sketches an alternative method beside the DVCS experiment 
in studying the $D$-term, pressure and shear forces inside the proton.
More precise data at JLab and future EIC in US \cite{Accardi:2012qut,AbdulKhalek:2021gbh} and China \cite{Chen:2018wyz,Chen:2020ijn,Anderle:2021wcy} 
are of importance for probing the $D$-term form factor and enhancing our understanding of the QCD structure of the proton.

\begin{acknowledgments}
We thank Prof. Fan WANG for the fruitful discussions and the suggestions.
This work is supported by the Strategic Priority Research Program of Chinese Academy of Sciences
under the Grant NO. XDB34030301.
\end{acknowledgments}

\bibliographystyle{apsrev4-1}
\bibliography{refs}

\end{document}